\documentclass[preprint,flushrt]{aastex}
\usepackage{natbib}
\usepackage{lineno}
\shorttitle{First {\it 3C454.3 giant flare}}
\shortauthors{Abdo et al.}

\title{Fermi Gamma-ray Space Telescope Observations of Recent Gamma-ray Outbursts from 3C~454.3}
\author{
M.~Ackermann\altaffilmark{1}, 
M.~Ajello\altaffilmark{1}, 
L.~Baldini\altaffilmark{2}, 
J.~Ballet\altaffilmark{3}, 
G.~Barbiellini\altaffilmark{4,5}, 
D.~Bastieri\altaffilmark{6,7}, 
K.~Bechtol\altaffilmark{1}, 
R.~Bellazzini\altaffilmark{2}, 
B.~Berenji\altaffilmark{1}, 
R.~D.~Blandford\altaffilmark{1}
% ...
}
\altaffiltext{1}{W. W. Hansen Experimental Physics Laboratory, Kavli Institute for Particle Astrophysics and Cosmology, Department of Physics and SLAC National Accelerator Laboratory, Stanford University, Stanford, CA 94305, USA}
\altaffiltext{2}{Istituto Nazionale di Fisica Nucleare, Sezione di Pisa, I-56127 Pisa, Italy}
\altaffiltext{3}{Laboratoire AIM, CEA-IRFU/CNRS/Universit\'e Paris Diderot, Service d'Astrophysique, CEA Saclay, 91191 Gif sur Yvette, France}
\altaffiltext{4}{Istituto Nazionale di Fisica Nucleare, Sezione di Trieste, I-34127 Trieste, Italy}
\altaffiltext{5}{Dipartimento di Fisica, Universit\`a di Trieste, I-34127 Trieste, Italy}
\altaffiltext{6}{Istituto Nazionale di Fisica Nucleare, Sezione di Padova, I-35131 Padova, Italy}
\altaffiltext{7}{Dipartimento di Fisica ``G. Galilei", Universit\`a di Padova, I-35131 Padova, Italy}
% ...
\begin{abstract}
The flat spectrum radio quasar  3C~454.3 underwent an extraordinary  outburst in December 2009 when it became the brightest $\gamma$-ray source in the sky for over one week. Its daily flux measured with the {\sl Fermi} Large Area Telescope at photon energies $E>100$ MeV reached $F_{100}  = 22\pm 1 \times 10^{-6}$ ph cm$^{-2}$ s$^{-1}$,  representing the highest daily flux of any blazar ever recorded in high-energy $\gamma$-rays. It again became the brightest source in the sky in 2010 April, triggering a pointed-mode observation by Fermi.  The  correlated  $\gamma$-ray temporal and spectral properties during these exceptional events are presented and discussed. The main results show flux variability over time scales less than 3 h and very mild spectral variability with an indication of  gradual hardening preceding major flares. No consistent loop pattern emerged in the $\gamma$-ray spectral index vs flux  plane.   A minimum Doppler factor of $\approx 15$ is derived, and the maximum energy of a photon from 3C~454.3 is $\approx$ 20 GeV. The spectral break at a few GeV is inconsistent with Klein-Nishina softening from power-law electrons scattering Ly$\alpha$ line radiation, and a break in the underlying electron spectrum in blazar leptonic models is implied. 
\end{abstract}
\begin{document}
\section{Introduction}
The radio source 3C~454.3  is a well-known flat spectrum radio quasar (FSRQ) at redshift $z=0.859$. It entered a bright phase starting in 2000, and has shown remarkable activity in the past decade. It underwent major outbursts in 2005, reaching an R-band magnitude of $12.0$ and the largest apparent optical luminosity ever recorded from a blazar \citep{Fuh06,Vil06,Gio06}. It also underwent major outbursts in 2007 \citep{Ghi07,Ver09} and 2008 \citep{Ver10,Jor10}.

First observations of 3C 454.3 with the Fermi Large Area Telescope (LAT) 
began in July 2008 during Fermi's commissioning period, 
when the source was found at a high flux state with 
$F_{E >100{\rm~ MeV}}\cong 10 \times 10^{-6}$\, ph \,cm$^{-2}$\,s$^{-1}$ \citep{Abdo_3C}. During this time, most observations were carried out in pointed mode with 3C 454.3 being close to the edge of the field of view ($\simeq 55^{\circ}$ off-axis), which did not allow for detailed spectral analysis during the brightest stage of the outburst. Observations in the decay stage, performed in survey mode, revealed a timescale less than 2 days for the flux to decline by a factor of 2. The spectrum showed a spectral break around 2 GeV with a spectral steepening from $\Gamma_1=2.3$ to $\Gamma_2=3.5$. Such a break has now been found  to be a common feature in bright FSRQs and in some low-synchrotron peaked BL Lacs as well \citep{spec_an}. Based on weekly light curves,  a very moderate ``harder when brighter'' effect has also been observed, with the photon spectral index obtained with a single power-law fit varying by less than 0.3 for flux ratios varying by $>$7 \citep{spec_an}. The source is listed as 1FGL J2253.9+1608 in the First LAT AGN catalog \citep{1LAC}. 

The continuous monitoring by the Fermi LAT showed that the source activity  faded continuously in early 2009 and then rose back up from June onwards. It  underwent an exceptional outburst in November 2009 - January 2010  when it became the brightest $\gamma$-ray source in the sky for over a week, reaching a record daily flux level in the GeV band as seen by the LAT and AGILE \citep{ATEL2322,Atel2326,Atel2328,Str10}. At the same time it also showed strong activity at optical frequencies \citep{Atel2325,Atel2332,Atel2333}, in the Swift/XRT and BAT bands  \citep{Atel2329,Atel2330}, and in the INTEGRAL/IBIS band \citep{Atel2344}. \cite{Tav10}  claimed $\gamma$-ray variability on timescales of a few hours from the LAT data and discussed its implications on the size and distance to the black hole of the emitting region, while \cite{Bon10} studied the single-day broad-band SEDs in the flare and in a lower state and found  that a single-zone external Compton plus synchrotron self-Compton model adequately fitted the data. However, with a similar analysis, \cite{Pac10} came to the conclusion that an additional component was required to fit the data.  The source remained active afterwards with a slowly decaying flux around 2$\times$10$^{-6}$ ph cm$^{-2}$ s$^{-1}$ until early April 2010, when it brightened up again to a flux level of $\approx 16 \times$ 10$^{-6}$ ph cm$^{-2}$ s$^{-1}$, prompting the first Fermi-LAT target-of-opportunity (ToO) pointed observation beginning on 2010 April 5 lasting for 200 ks.    

These two major events offer a unique opportunity to probe intraday variability 
and the associated spectral changes in the $\gamma$-ray band, which is the focus of this paper.  Particular attention is paid to the correlated spectral/flux variations on different time scales, both in terms of spectral hardness and position of the spectral break, which have not been investigated in detail before.
 In Section 2, observations and analysis of data from 3C~454.3 from August 2009 through
April 2010 is presented. Section 3 gives results of the analysis, and Section 4 provides interpretation.
We summarize in Section 5.
A flat $\Lambda$CDM cosmology with H$_0$ = 71 km s$^{-1}$ Mpc$^{-1}$, $\Omega_m$ = 0.27, $\Omega_\Lambda$=0.73 
is used in this paper, implying a luminosity distance $d_L = 1.69\times 10^{28}$ cm to 3C 454.3.

\section{Observations and analysis}

The {\it Fermi}-LAT is a pair-conversion $\gamma$-ray telescope sensitive to photon  energies greater than 20 MeV.
In its nominal scanning mode, it 
surveys the whole sky every 3 hours with a field of view of about 2.4 sr \citep{Atw09}. 
The data presented in this paper (restricted to the 100 MeV-200 GeV range) were collected from 2009 August 27 to 2010 April 21 in survey mode, except for a 200 ks period starting on  April 5 at 19:38 UT when the pointed mode was used, resulting in a gain of rate of accumulation of exposure by about a  factor of 3.5 over survey mode. During pointed mode, the source direction was offset by $10^\circ$ with respect to the LAT axis  in order to limit adverse effects related to gaps in the detector that can affect on-axis photons. Over 3000 photons with $E>100$ MeV were collected in pointed mode.  To  minimize systematics, only photons with energies greater than 100 MeV were  considered in this analysis. In order to avoid contamination from Earth limb  gamma-rays, a selection on events with zenith angle $<105^{\circ}$ was applied.  This analysis was performed with the standard analysis tool {\it gtlike}, part  of the Fermi-LAT ScienceTools software package (version v9r15p5). The P6\_V3\_DIFFUSE  set of  instrument response functions was used. This set includes a  correction  for the average reduction of the effective area due to pile-up effects as fewer photon events pass the rejection cuts. This correction is sufficient for integration times longer than a day. The residual energy and trigger-rate dependent acceptance variations present for shorter times as established with Vela and Galactic diffuse emission data  (typically amounting to less than to 10\% in the considered periods), have not been corrected for in this analysis. 

Photons were selected in a circular region of interest (ROI) 10$^\circ$ in radius, centered at the position of 3C 454.3.  The isotropic background, including the sum of  residual instrumental background and extragalactic diffuse $\gamma$-ray background, was modeled by fitting this component at high galactic latitude (file provided with ScienceTools).  The Galactic diffuse emission model version ``gll\_iem\_v02.fit," was used, with  both flux and photon spectral index left free in the fit (the Galactic longitude and latitude of 3C 454.3 is $86.1^\circ$ and $-38.2^\circ$, respectively).  All point sources lying within the ROI and a surrounding  5$^\circ$-wide annulus with a flux greater than about 1\% of the quiescent level of 3C 454.3 were modeled in the fit with single power-law  distributions.

Although the actual spectral shape is better reproduced by a broken power law, the spectral variations were investigated using the photon indices resulting from single power-law (PL) fits, which provided the best statistical accuracy. All light curves were produced using fluxes derived with power-law fits.

Different analyses were performed by fitting the spectra with  various models over the whole energy range covered by the LAT at  $E> 100$ MeV: a broken power law 
$N(E)=N_0 (E/E_{break})^{-\Gamma_{i}}$, with $i=1$ if $E<  E_{break}$ and $i=2$ if  $E >E_{break}$,  a log parabola function \\ (\mbox{$N(E)=N_0 (E/E_{p})^{-\alpha-\beta log(E/E_p)}$} where $E_p$ is fixed at 1 GeV) and  a power law with exponential cutoff function  (\mbox{$N(E)=N_0 (E/E_{0})^{-\Gamma}\exp(-E/E_{cutoff})$}), and with a PL model over equally spaced logarithmic energy bins with the spectral index kept constant and equal to the value fitted over the whole range.

In  case of fits with broken power law (BPL) models, the break energy  (E$_{break}$), which separates the photon energy ranges where different photon indices $\Gamma_1$ (for $E<E_{break}$) and  $\Gamma_2$ (for $E>E_{break}$) apply, could not be obtained directly from the fit because of convergence problems due to the non-smooth character of the BPL function at the break energy. It was instead computed from  a log-likelihood profile fitting procedure, with a statistical uncertainty  corresponding to a difference of $-2\Delta L$=1 in the log-likelihood function $L$ with  respect to its minimum. 

In order to minimize spurious correlations between flux and spectral index, the fluxes $F_{\rm E>E_1}$  were also computed above the photon energy $E_1$ where the correlation between integrated flux and index is minimal. This energy was derived from a power-law fit of the form \\ \mbox{$N(E)=F_{\rm E>E_0}(\Gamma-1)\:E^{-\Gamma}/E_0^{-\Gamma+1}$} (with $E_0$=100 MeV).  Since $F_{\rm E>E_1}=F_{\rm E>E_0} (E_1/E_0)^{-\Gamma+1}$, minimizing  $\Delta F_{\rm E>E_1}/F_{\rm E>E_1}$ with respect to $E_1$ yields  $\ln(E_1/E_0)=C_{F\Gamma}/(F_{\rm E>E_0}\:C_{\Gamma \Gamma}$), where $C_{F\Gamma}$ and $C_{\Gamma \Gamma}$ are terms of the covariance matrix returned by the fit. In the 3C 454.3 case, $E_1$ has been found to be 163 MeV over the ToO time range. The same value has been used for the other time periods as well.
 
The estimated systematic uncertainty on the flux is 10\% at 100 MeV, 5\% at 500  MeV and 20\% at 10 GeV. The energy resolution is better than 10\% over the range  of measured E$_{break}$.

\section{Results}

\begin{figure}
\epsscale{0.80}
\plotone{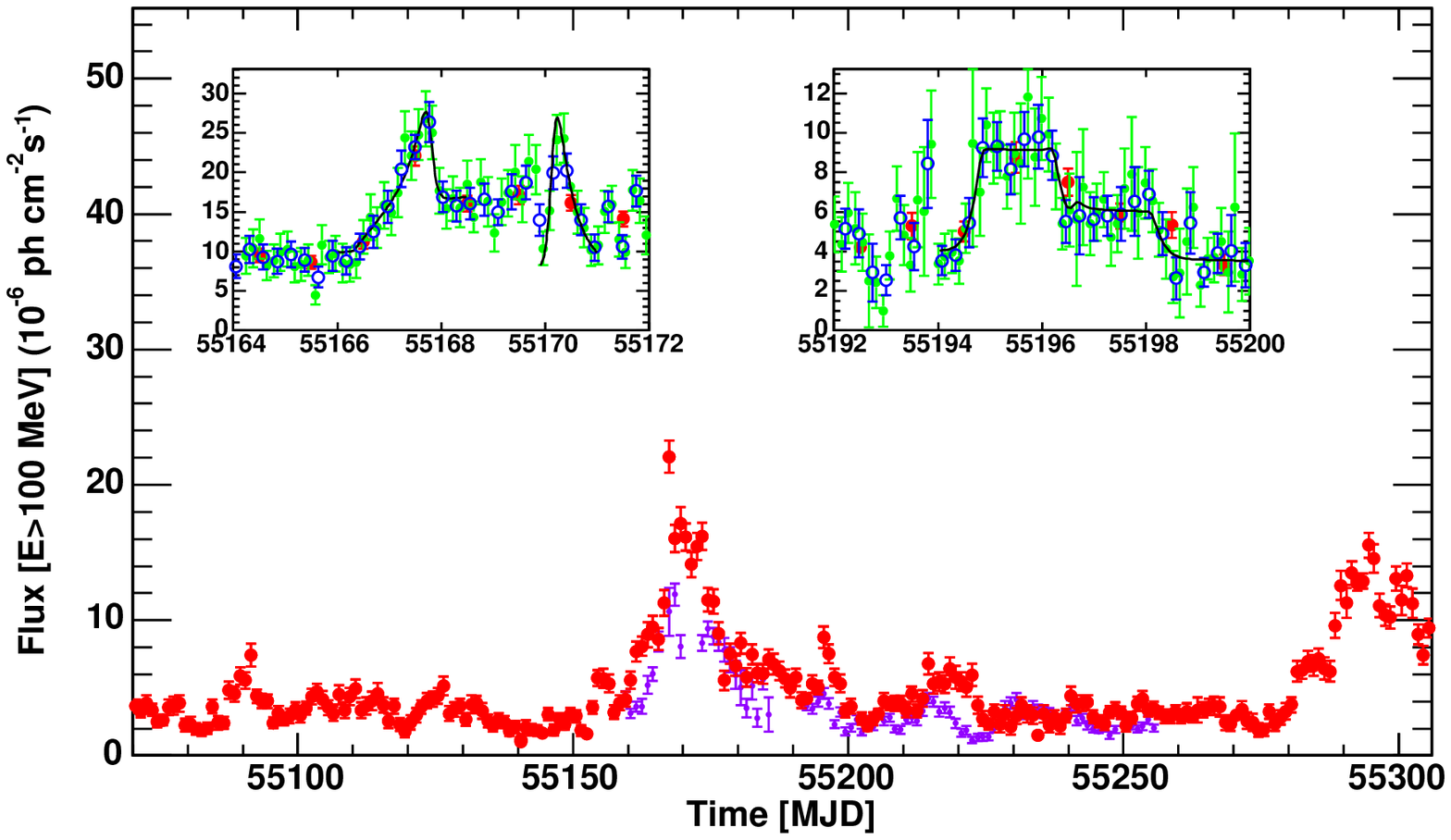}
\caption{The light curve of the flux of 3C~454.3 in the
$100\,$MeV -- $200\,$GeV band (red). The light curve of the July - August 2008 flare, shifted by 511 days, is shown for comparison (magenta). The insets show blow-ups of the two periods when the
 largest relative flux increases took place.  The red, blue, and green data points in the insets correspond to daily, 6 hr, and 3 hr averaged fluxes, respectively. The fit results discussed in the text are displayed as solid curves. }
\label{fig:light_curve}
\end{figure}

Figure \ref{fig:light_curve} displays the daily light curves (red points) from MJD 55070 to 55307  (2009 August 27 to 2010 April 21) for fluxes above 100 MeV. The periods showing the  fastest flux variations during the December flare, with fluxes  changing by more than a factor of 2 in amplitude, are enlarged in the insets, with the $E>100$ MeV fluxes averaged over a daily, 6-hour, and 3-hour binning shown by the red, open blue and green data points, respectively. The error bars are statistical only. Three flares displaying a flux variation greater than a factor of 2 over less than a day (MJD 55167, 55170 and 55195)  have been studied extensively during this period. These flares exhibit very different profiles and degrees of symmetry. The first two flares were fitted with the function \citep{Abdo_var}
\begin{equation} 
F=2F_0(e^{(t_0-t)/T_r}+e^{(t-t_0)/T_f})^{-1}+F_{bgd}(t), 
\label{fit_function}
\end{equation}
where $T_r$ and $T_f$ are the rising and falling times, $F_0$  is the flare flux amplitude and $F_{bgd}(t)$ is a (slowly varying) background flux, while for the third one a constant plateau flux was allowed between the rising and falling phases. The characteristic flare duration can be estimated as $T_r$+$T_f$. We obtained  $T_r$=0.37 d and  $T_f$=0.06 d for the MJD 55167 (Dec.\ 2) flare, and  $T_r$=0.07 d, $T_f$=0.26 d for the MJD 55170 (Dec.\ 5) flare. The MJD 55195 flare, which is less prominent, shows a rise time of $\simeq 0.16$ d.  In the two early flares, the flux variation $F_0$ was greater than  10$^{-5}$ ph$(>100$ MeV) cm$^{-2}$ s$^{-1}$, and statistically significant factor-of-2 variations take place on timescales as short as 3 hours. 

In  Figure \ref{fig:light_curve}, the light curve of the July-August 2008 outburst \citep[shifted by 511 days; see][]{Abdo_3C} is shown for comparison. The resemblance of the two light curves is notable , although the estimated fluences are different by 40\%. The brightest periods of the outbursts last for about 10 days, and are then followed by a long tail of fairly high activity upon which are superimposed minor flares lasting for a few days. Although the outburst in April 2010 is longer, lasting for $>$ 30 days, it exhibits similar patterns as the December 2009 outburst, where its maximum level is preceded by an intermediate elevated flux lasting for about 5 days. This feature could serve as an alert for an imminent surge in flux. 
 
\begin{figure}[t]
\centering
\resizebox{13cm}{!}{\rotatebox[]{0}{\includegraphics{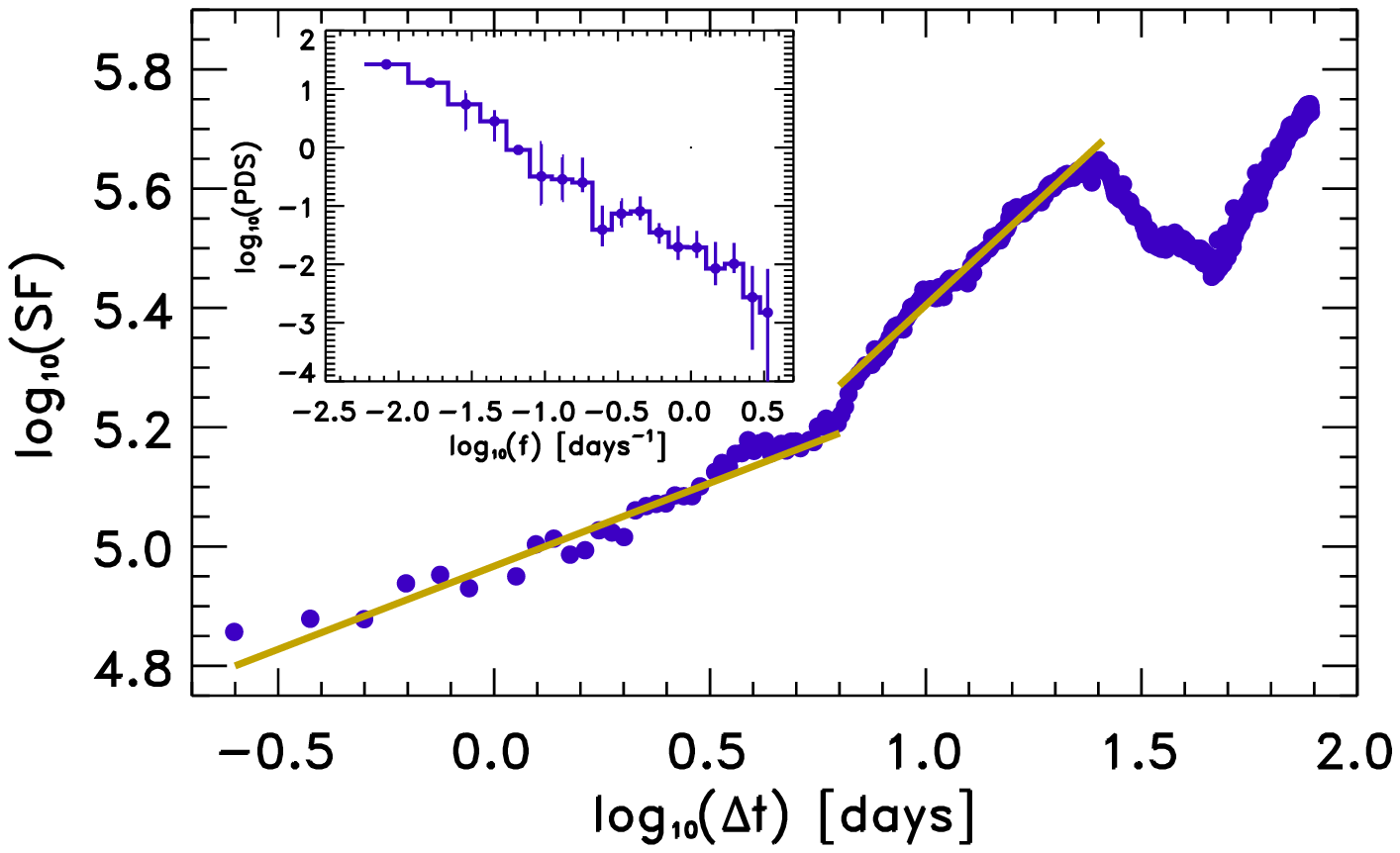}}}\\
\resizebox{13cm}{!}{\rotatebox[]{0}{\includegraphics{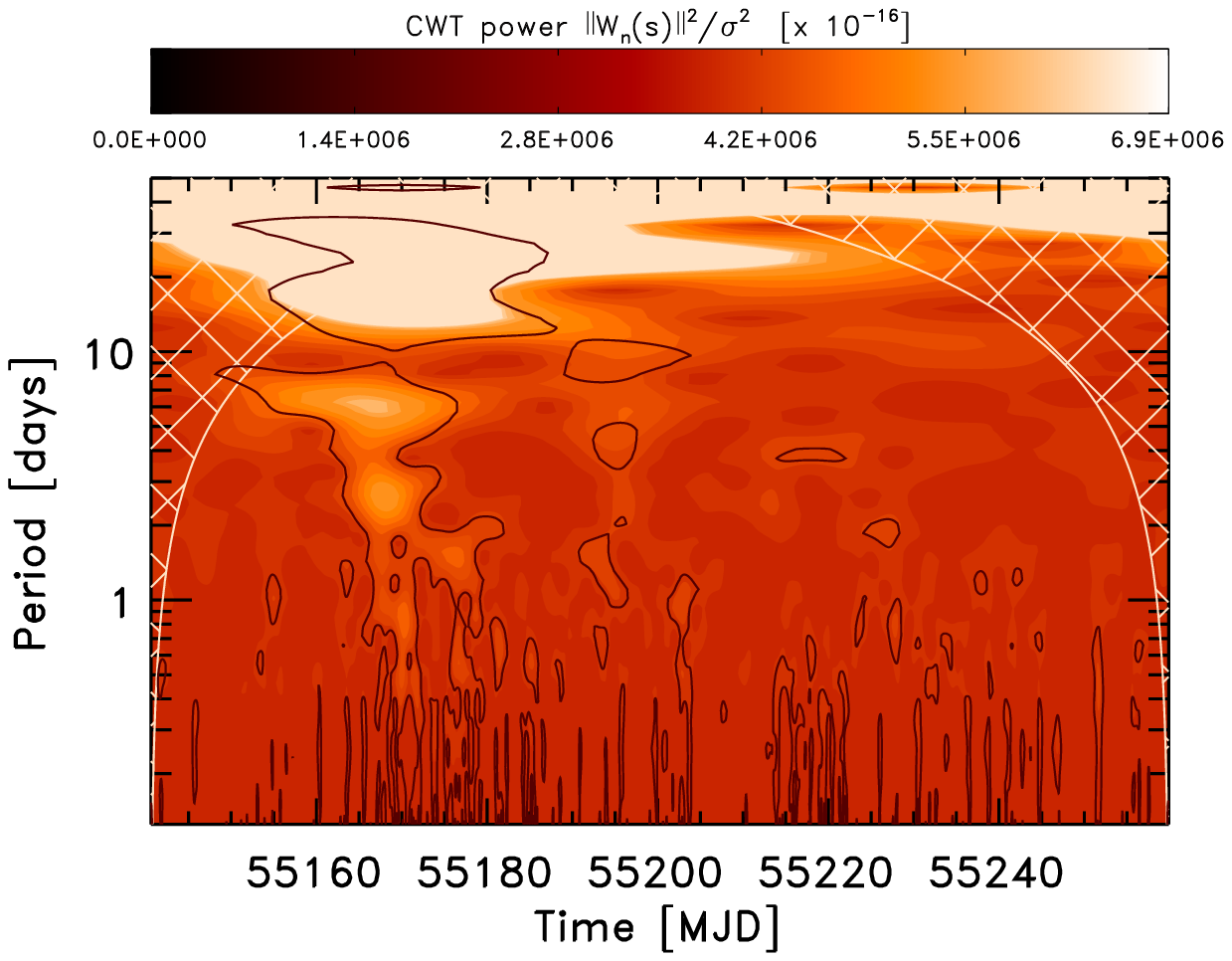}}}
\caption{Top panel: structure function of the 3h-bin flux light curve for the period 2009 Nov. 5 - 2010 Mar. 4 and corresponding power density spectrum (inset). Bottom panel: plane contour plot of the continuous Morlet wavelet transform power density for the same light curve. Thick black contours are the 90\% confidence levels of true signal features against white/red noise background, and cross-hatched regions represent the ``cone of influence'', where spurious edge effects become important.} \label{figure_sf_wavelet}
\end{figure}

The power density spectrum (PDS) of the  2009 Nov. 5 - 2010 March 4 3h-light curve has been calculated in a similar way as that reported in \cite{Abdo_var} for the  11-month light curve (3-day
bin),  normalized to fractional variance per frequency unit ($f$) and white-noise subtracted. This PDS indicates a power law power density, $1/f^{a}$ with index $a= 1.50\pm 0.06$, i.e., intermediate  between flickering (red noise) and shot noise (driven by
Brownian processes). Such result confirms over a wider frequency range the value found with the first 11 months of  data. In Figure \ref{figure_sf_wavelet} the first order structure function (SF), the PDS and the Morlet continuous wavelet transform (CWT) of the continuous, unprecedented-resolution, $\gamma$-ray light curve
of 2009 Nov. - 2010 March  are reported. A break around 6.5 days is suggested by
the SF analysis, the power index slopes being $a=1.29\pm0.10$ between 3 hours and 6.5 days, and $a=1.64\pm0.10$ between 6.5 and about
26 days, while at longer lags spurious drops due to the finite range become apparent. The PDS analysis confirms these values ($a=1.40 \pm 0.19$ at high frequency and $a=1.56 \pm 0.18$ at low frequency). The temporal behavior of 3C~454.3 is therefore showing $1/f^{1.5}$ universality
from 3-hours to 11-months timescales. The 6.5-day break, consistently confirmed by the PSD and more clearly depicted by the SF, represents a steepening towards longer lags (flattening towards higher frequencies) and does not necessarily imply a local characteristic time scale, as it could simply be the point where two PDS components with different slopes are equally strong. Gamma-ray variability in 3C 454.3 can be seen as a short term realization of a stochastic mechanism, where structures that are resolved in shorter observations are simply averaged out in long observations, and where big outbursts correspond to statistical tails of the same process.
                                                                                
The CWT in Figure \ref{figure_sf_wavelet} provides a local and detailed time series analysis, through the two-dimensional energy density function (modulus
of the transform, filled color contour) computed using a Morlet waveform,  providing the best tradeoff between localization and period/frequency resolution.  For timescales below 1 day no local peaks are found, although some marginal features in this time range are found during the outburst state. The big outburst of December 2009 is, in fact, localized and decomposed in a chain of well-defined power CWT peaks. The 6.5 days timescale is confirmed by the major peak out of the cone of influence (localized at about MJD 55166, i.e,  the onset of the outburst, MJD $\sim$ 55166.2 - 55172). A second energetic peak in this period is found at about 2.5 days manifesting another dominating timescale during the outburst, while in the second period of the outburst two minor power peaks at about 19h and 1.3 days are also visible. Based on this CWT local analysis there is no evidence for structure on time scales shorter than about 12 hours but  shorter time scales close to  the sampling scale cannot be ruled out.                                                                              

\begin{figure}
\epsscale{0.80}
\plotone{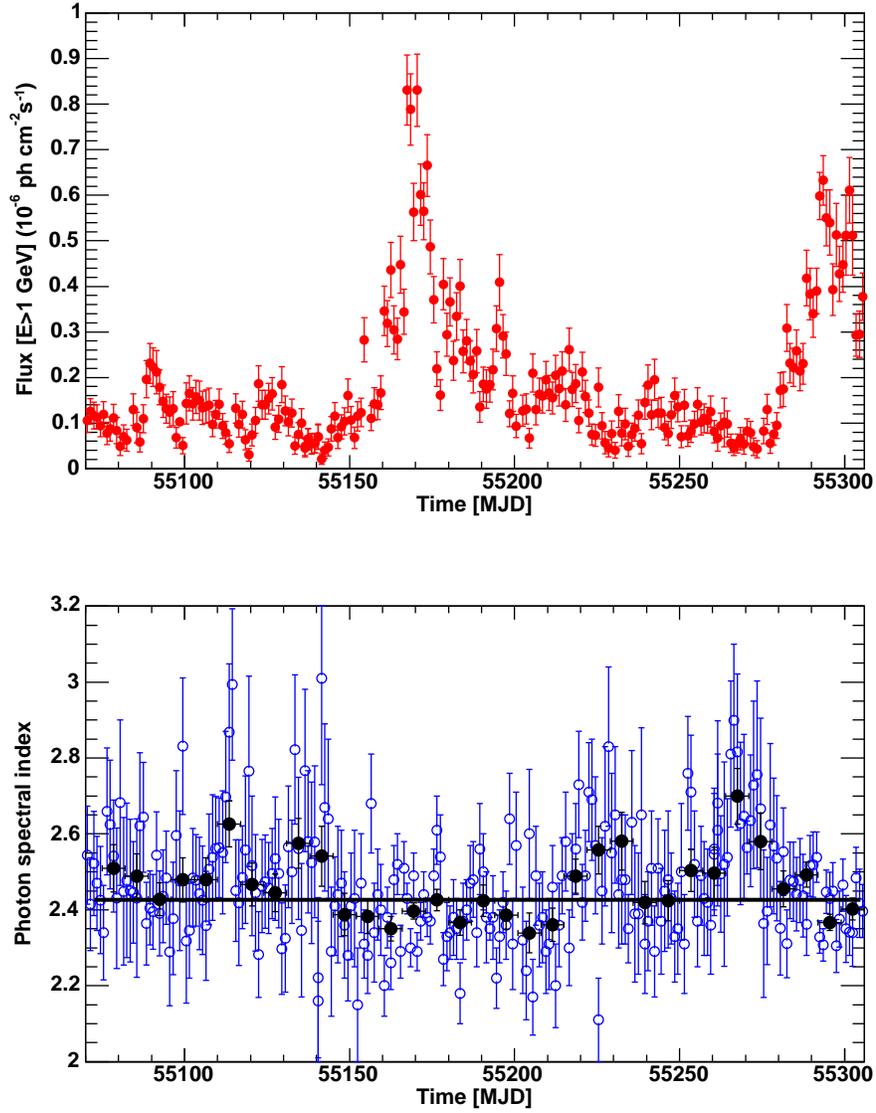}
\caption{Top:  Light curve of the flux in the $1\,$GeV -- $200\,$GeV band. Bottom: Variation of the daily (blue points) and weekly (black points) photon spectral index derived from a power-law fit. The black line depicts the mean  weekly spectral index. }
\label{fig:light_curve_a}
\end{figure}

The 1 GeV daily light curve, shown in Fig. \ref{fig:light_curve_a} top, 
 closely resembles the 100 MeV light curve, hinting at little spectral variability. This behavior is confirmed by the very limited variation of the photon spectral index measured at $ E > E_1 = 163$ MeV, displayed in  Figure \ref{fig:light_curve_a} bottom  by the daily average photon index (open blue symbols) as well as the weekly-averaged ones (solid black points). The near constancy of the spectrum is in accord with the results found from the July 2008 flare and the first 6-months of LAT data \citep{Abdo_3C,spec_an}.


The variation of the amplitude of the weekly photon indices is only $\Delta \Gamma$=0.35 (varying between 2.35 and 2.7) during the  period under consideration, but the variation is statistically significant. Comparing the weekly photon indices to their weighted average returns a reduced $\chi^2$ =86.4/32, corresponding to a 10$^{-7}$ probability for a nonvariable signal. Particularly notable is a slight softening of the spectrum during the periods of lowest fluxes, particularly during the periods of MJD 55221-55235 and 55264-55278. There is also a suggestion that a progressive hardening over several weeks precedes a major outburst,
but more such events will be needed to establish whether this behavior is typical.

The correlation between daily spectral index and flux was further
investigated by computing a correlation function  for four segments 
of the November 2009 - April 2010 time range. In the first three segments (Dec 2009 flare,
flare decay and inter flare region) we find evidence for a lag
such that spectral index variations lead the flux by about 5 days.
In contrast the correlation during the April 2010 flare is weak
and if anything shows a lag in the opposite sense. Although
the significance of the lag in these data is only about 2 $\sigma$
we do see the same 5-day lag also in the data from the 2008 flare.

\begin{figure}
\epsscale{0.7}
\plotone{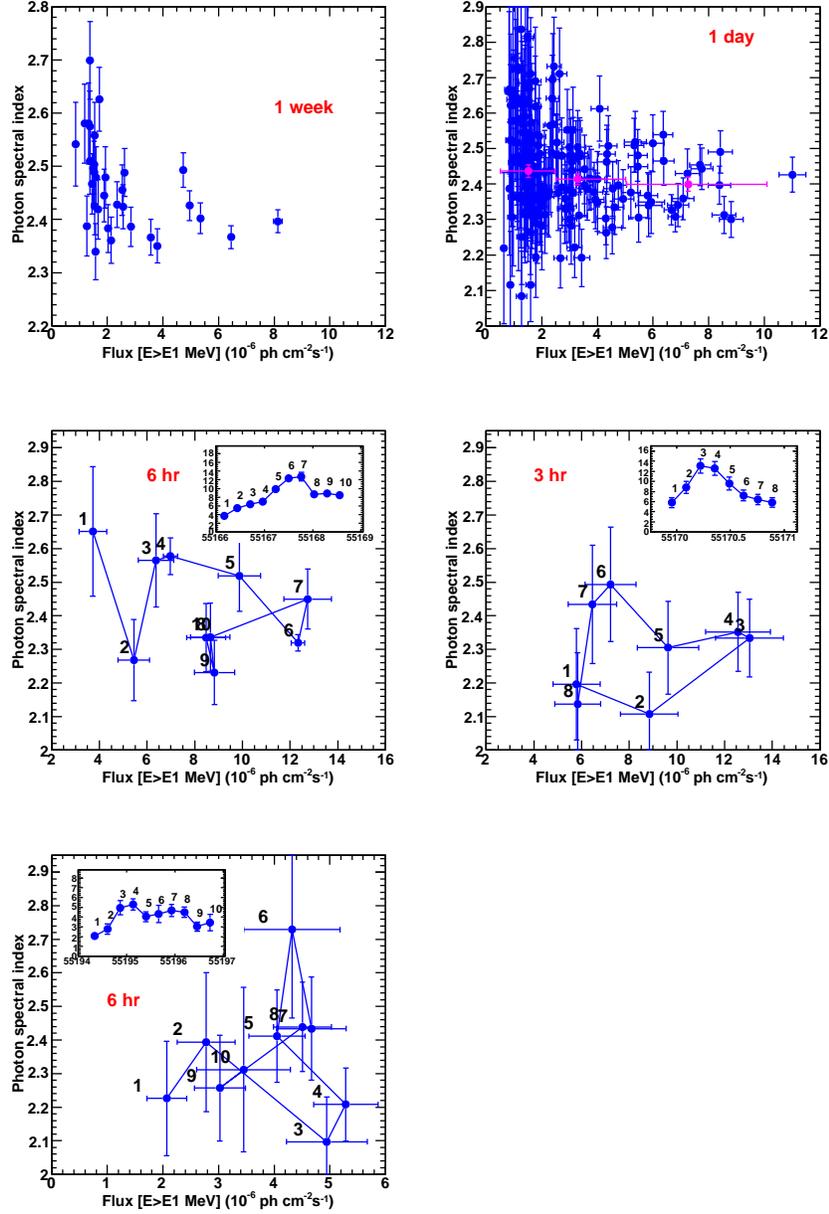}
\caption{Photon spectral index vs.\ flux measured on a weekly (upper left) and daily (upper right) basis are shown by blue points. 
The averages over periods with daily fluxes $F_{E_1}< 2.5$, $2.5<F_{E_1}<5$, $5<F_{E_1}<10$, where $F_{E_1}$ is the photon flux above $E_1$ in units of $10^{-6}$ ph cm$^{-2}$ s$^{-1}$, are shown by the magenta data points in the upper right panel. Middle left, middle right, and bottom panels show photon index vs.\ flux for the MJD 55167, MJD 55170 and MJD 55195 flares, respectively. The labels refer to the different times of the light curve as shown in the insets.
\label{fig:flux_index}}

\end{figure}

   Figure \ref{fig:flux_index} top left (right) presents the weekly- (daily-) averaged power-law photon spectral index vs.\ flux above $E_1$. A weak ``harder when brighter" effect shows up for weekly bins, but is almost washed out when considering daily bins.  To make the trend clearer, photons were sorted in 3 daily-flux bins ($F_{E_1}<$2.5, $2.5<F_{E_1}<5$, and $5<F_{E_1}<10$, where $F_{E_1}$ is the photon flux above $E_1$ in units of $10^{-6}$ ph cm$^{-2}$ s$^{-1}$) and the analysis was repeated with the resulting photon files. The result, displayed as magenta points in the top right panel of Figure \ref{fig:flux_index}, still exhibits a slight harder when brighter effect. The data point at $F_{E_1}\cong 11$ corresponds to the December 2$^{\rm nd}$ flare and is consistent with the trend observed at lower flux. The three other panels in Figure \ref{fig:flux_index} show the photon spectral index vs.\ flux patterns for the three rapid flares, obtained either with a 6-hour binning (middle left and  bottom left) or a 3-hour binning (middle right). The light curves at $E>E_1$ (not above 100 MeV, as in Fig.\ 1) calculated at the time of the main flaring episodes are displayed in the insets of Figure \ref{fig:light_curve}. Despite the flux reaching the highest values for a non-GRB cosmic source, the statistical significance of these patterns is marginal, except for the MJD 55167 flare.  Instead of a well-defined, universal pattern, a  variety  of patterns is found. The MJD 55167 flare is associated with a clockwise pattern, with a flux-doubling accompanied by an essentially constant (or weakly harder) photon spectral index.  The reduced $\chi^2$ for a constant fit of the photon spectral index is 28.6/9. The pattern is somewhat indicative of a ``hard-lag'' effect possibly reflecting particle acceleration \citep{Kir99}.  
The short MJD 55170 flare shows an indication for a counterclockwise loop ($\chi^2_r$ = 6.0/7). The loop diagram for this flare on 3-hour timescales reveals the simplest ordered pattern, in this case a softening followed by a flux
increase which then decays in flux and hardness. This could reflect the
underlying timescale on which coherent variability takes place in 3C
454.3. Because this pattern does not recur in other flaring episodes, no strong
conclusions can be made, however. The MJD 55195 flare shows some  softening during the plateau (points 3-6) following the flux rise as expected from a cooling behavior, but the significance is low ($\chi^2_r$ = 8.6/9).  While it is difficult to draw any firm conclusion on acceleration and cooling from these patterns, the lack of strong spectral variability still provides clues to the underlying physical processes.  

\begin{figure}
\epsscale{.90}
\plotone{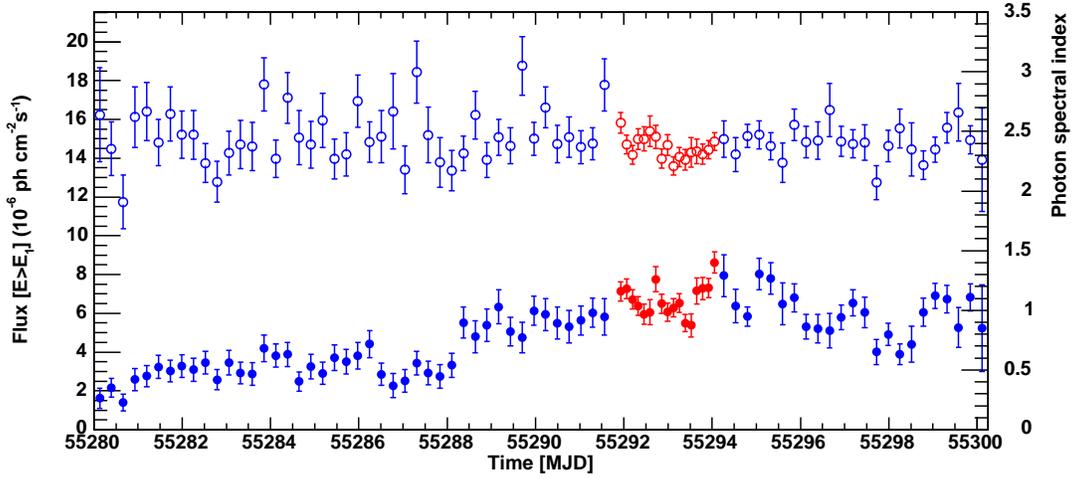}
\caption{Flux (filled data points; left-hand axis) and photon index (open data points; right-hand axis) as a function of time in the period surrounding the ToO pointing. Data collected in survey mode (6-hour binning) are in blue, those collected in pointed mode are in red (3-hour binning)}
\label{fig:light_curve_ToO}

\end{figure}

\begin{figure}
\epsscale{.7}
\plotone{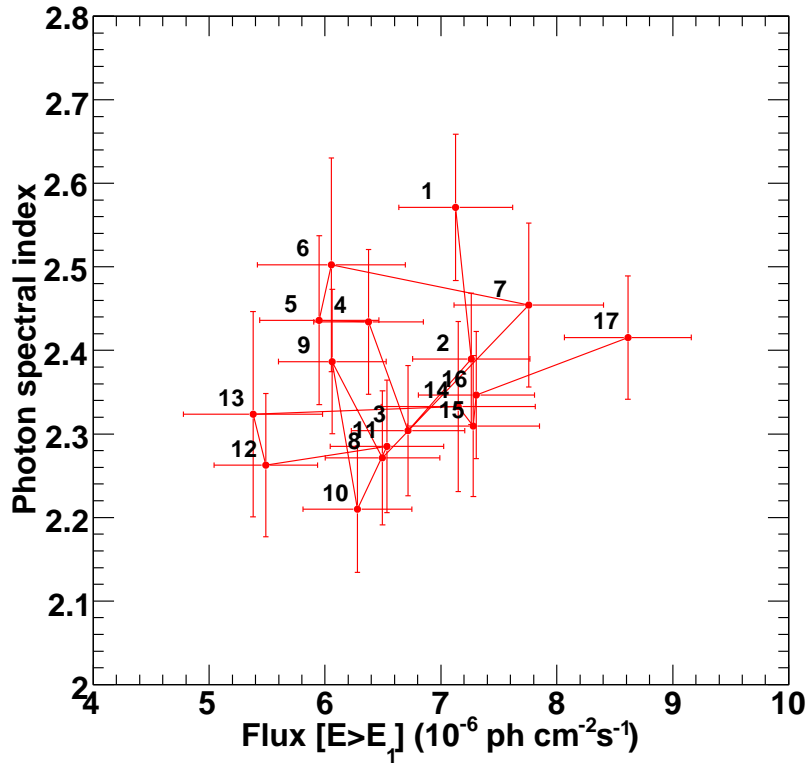}
\caption{Photon spectral index vs flux on 3-hour timescales during the time of the ToO pointing.
\label{fig:flux_index_ToO}}

\end{figure}

Figure \ref{fig:light_curve_ToO} shows the flux and photon spectral index as a function of time in the period around (blue symbols) and during (red symbols) the time of the target of opportunity when the Fermi LAT was in pointed mode (MJD 55291.82-55294.13). The binning is 6 hour and 3 hour for the survey and pointed modes respectively. As expected by the 3.5 fold increase in exposure per unit time during the ToO, the statistical accuracy in the measurement of  both parameters improves significantly. Although in a high state, the source was unfortunately fairly steady during this period. No indication for variability more rapid than that observed during the giant outburst is found during the ToO period, as already noted by \cite{Fos10}. The photon spectral index vs.\ flux above E$_1$ measured in 3-hour time bins is plotted in Figure \ref{fig:flux_index_ToO}. The reduced $\chi^2$ for a constant fit of the photon spectral index is 18.52/16, indicating that the data are consistent with a constant value. The correlation coefficient is 0.26.  
      
\begin{figure}
\epsscale{0.6}
\plotone{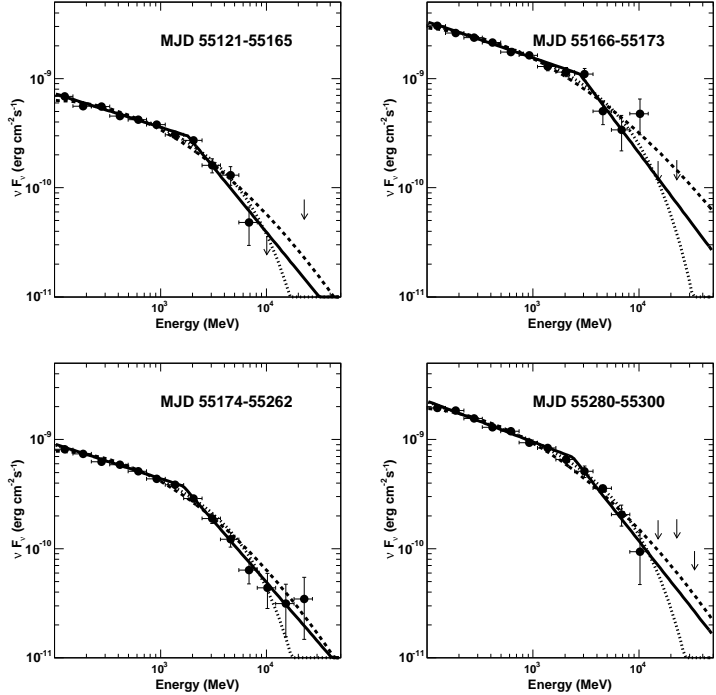}
\caption{ $\nu F_{\nu}$ distributions of the Fermi LAT data for different time periods, along with the fitted broken power law (solid), log parabola (dashed) and power law+exponential cutoff (dotted) functions.
\label{fig:nuFnu}}
\end{figure}

Figure \ref{fig:nuFnu} shows the integrated spectrum from Period 
1, MJD 55121-55165, Period 2, MJD 55166-55173 (week of the giant outburst), Period 3, MJD 55174-55262 and Period 4 MJD 55280-55300. The distributions have been fitted with a broken power law,  a log parabola function and  a power law with exponential cutoff function. The log parabola  gives a worse fit than the broken power law and the  power law with exponential cutoff functions, which  are difficult to discriminate for these periods. The fitted parameters are given in Table \ref{tab:funct} for the four periods.  

\begin{figure}
\epsscale{1.0}
\plottwo{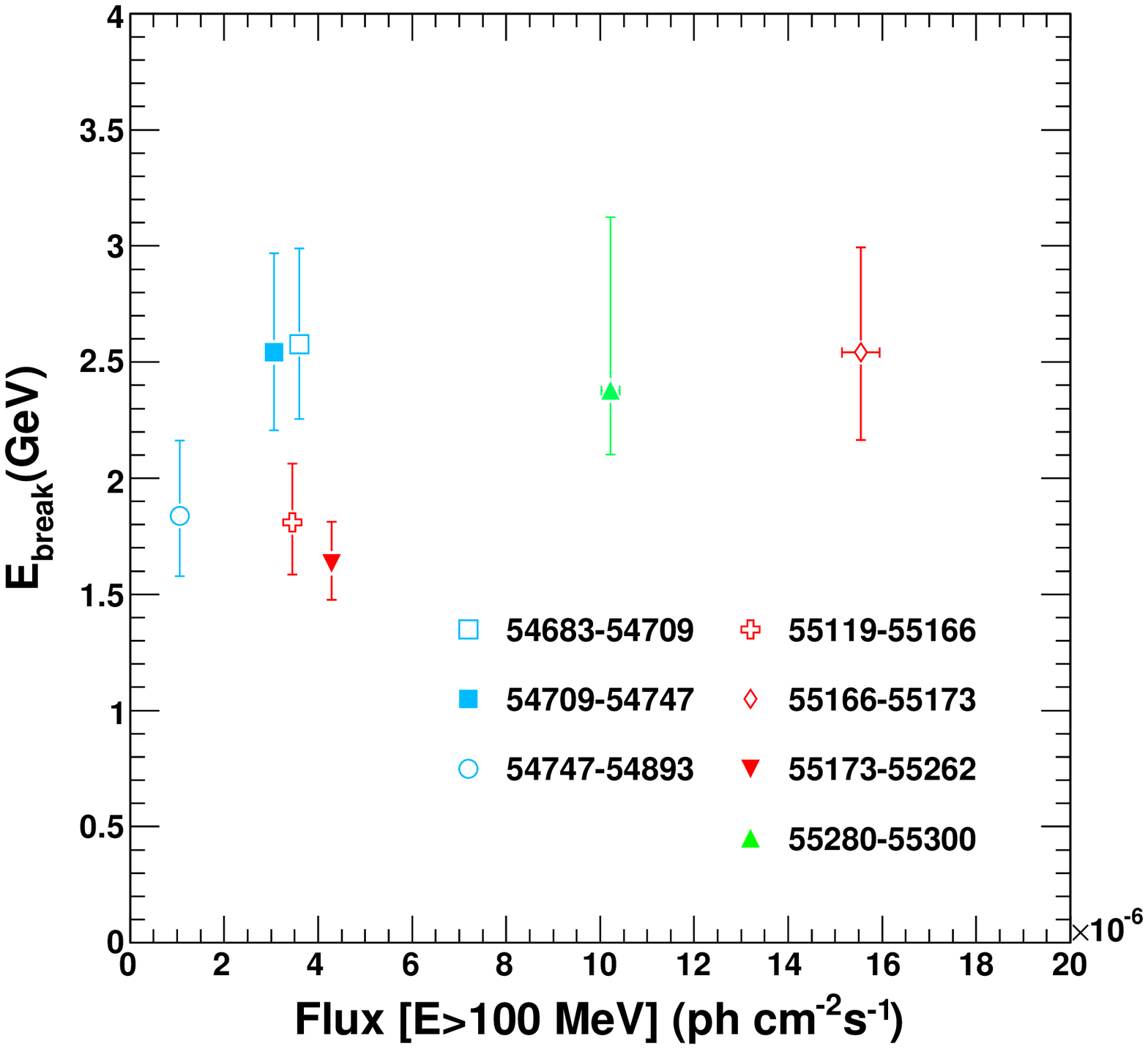}{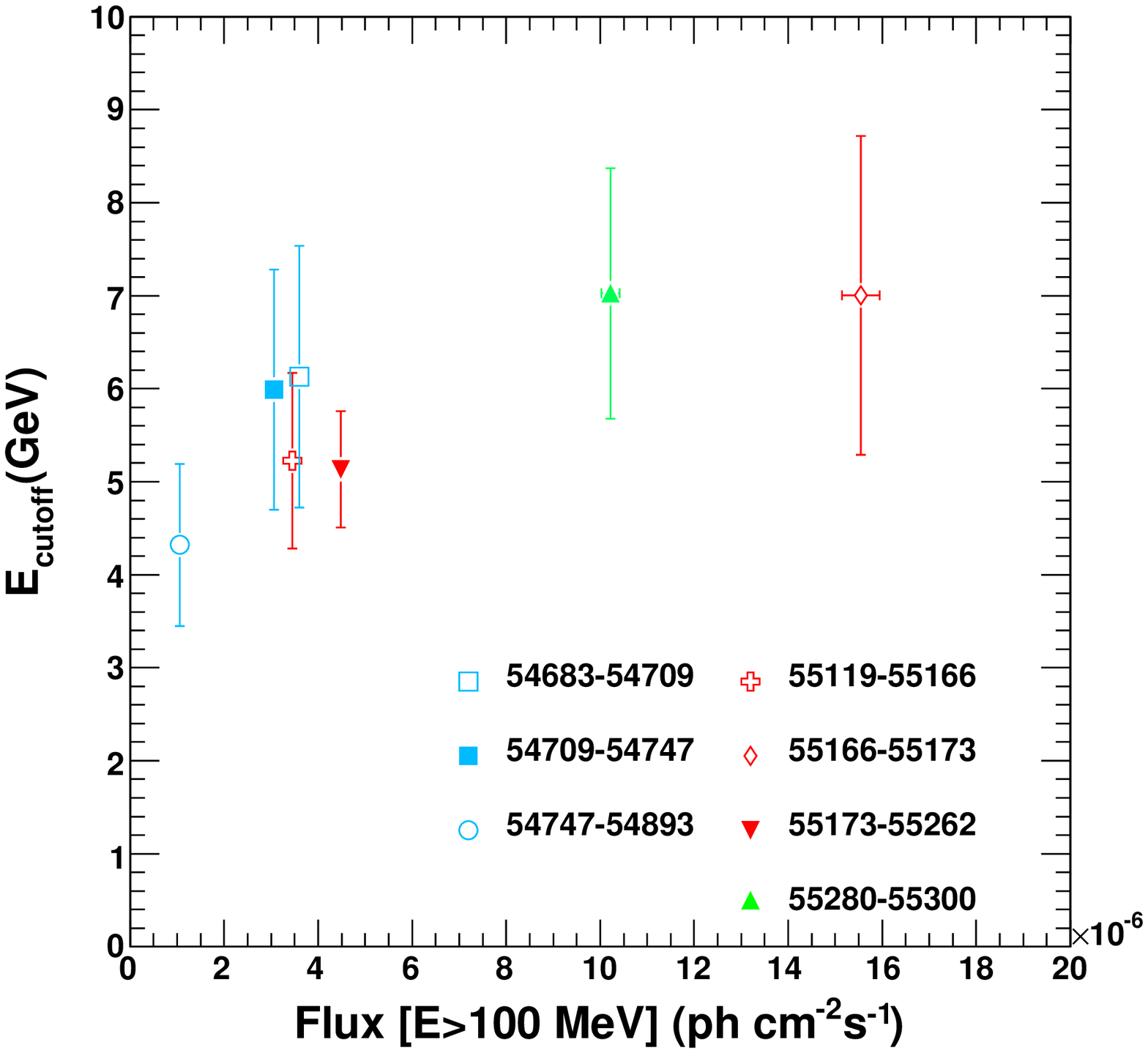}
\caption{Left: $E_{\rm break}$ vs flux for different time periods (given in MJD). Right: 
 $E_{\rm Cutoff}$ vs flux for the same time periods.
\label{fig:Eb_flux}}
\end{figure}

The variation of break energy (cutoff energy) with flux is displayed in the left (right) panel in Figure \ref{fig:Eb_flux} at different observing periods. No strong evolution of either the break energy or the cutoff energy is found, but there is some indication of a slight hardening with flux.  For a given flux, the position of the break energy is slightly different from that observed during the bright outburst in 2008, but all  E$_{break}$ (and E$_{cutoff}$)  are constant within a factor of $\approx 2$.  For the same periods \cite{Str10} found  fairly large spectral variability with the AGILE data, with a photon spectral index as low as 1.60$\pm$0.32 during big flares. This behavior is not confirmed by the present analysis.  

The maximum photon energy found within the 95\% containment radius from the location of 3C 454.3 during the period MJD 55140-55261 was a photon with $E= 20.7$ GeV at  MJD 55179.98, when $F_{E >100{\rm~ MeV}} = 6\times 10^{-6}$ ph cm$^{-2}$ s$^{-1}$ and the variability time $t_{var} \approx 1$ day. On December 2, when the average of $F_{E >100{\rm~ MeV}}$ over this day exceeded $\approx 20\times 10^{-6}$ ph cm$^{-2}$ s$^{-1}$, the highest energy was 8.5 GeV. As already noted, the flux variations during this flare were on timescales as short as a few hours or less. The energy of the most energetic photon  during periods of rapid flux changes provides the strongest constraint on the $\gamma\gamma$ opacity, as discussed below.

\section{Discussion}

A photon flux of $F_{E >100{\rm~ MeV}}  = 22\pm 1\times 10^{-6}$ ph cm$^{-2}$ s$^{-1}$ from 
3C 454.3 at $z = 0.859$
implies an apparent isotropic $\gamma$-ray luminosity above 100 MeV of $L_{\gamma}\cong 3.3\pm 0.15\times 10^{49}$ erg s$^{-1}$. This is $\sim 3$ times larger than the luminosity of the $z=1.839$ blazar PKS 1502+106 during its August 2008 flare \citep{2010ApJ...710..810A}, but still lower than the luminosity of PKS 1622-297 ($\sim 5.1\times 10^{49}\;{\rm erg\;s^{-1}}$ with the current cosmological model) during the 1995 flare \citep{Mat97}.
The time-averaged $\gamma$-ray luminosity (Table \ref{tab:funct})
measured with the Fermi LAT is $\approx 9\times 10^{48}$
and $\approx 1.6\times 10^{49}$ erg s$^{-1}$ for periods 1 and 2, respectively.
Estimates for the black-hole mass in 3C 454.3 range from $M_9\approx 1$ 
\citep{Bon10} to $M_9\approx 4$ \citep{Gu01}. 
To be radiating below the Eddington luminosity of 
1.26 $\times 10^{47}(M_9 M_\odot)$ erg s$^{-1}$ 
for a 10$^9 M_9$ Solar mass
black hole implies that the high-energy 
radiation is beamed into a jet with opening angle $\theta_j\lesssim0.1$

The light curves from Fig. \ref{fig:light_curve} show evidence for variability timescale of few hours and the model fitted to a very brief subflare at MJD 55170 implies a flux doubling timescale of $(\ln 4)\times T_r\sim 2.3\;{\rm h}$. However, if the bulk of the flare were characterized with such a short timescale, we would expect a very erratic light curve calculated with daily binning. Because the overall shape of the November/December 2009 flare is smooth, the dominating variability timescale is rather close to $t_{\rm var}\sim 1\;{\rm d}$. The short subflares may reflect existence within or outside the main emitting zone of geometrical substructures as compact as $R=\delta ct_{\rm var}/(1+z) \sim 3.3\times 10^{15}\delta_{25}t_{\rm var,2.3h}\;{\rm cm}\sim 22\delta_{25}t_{\rm var,2.3h}M_9R_{\rm g}$.

The minimum Doppler factor $\delta_{min} < \delta\equiv [\Gamma(1-\beta \cos \theta)$]$^{-1}$, where $\theta$ is 
the angle from the jet axis to the line of sight, is 
defined by the condition $\tau_{\gamma\gamma}(\epsilon_1)= 1$,
and can be estimated to $\approx 10$\% accuracy compared to results of 
more detailed numerical calculations through the expression
\begin{equation}
\delta_{min}\cong \left[\frac{\sigma_T d_L^2(1 + z)^2 f_{\hat{\epsilon}} \epsilon_1}{4t_{var}m_ec^4}\right]^{1/6}
\label{deltaD}
\end{equation}
\citep{Don95,Ack10}, 
where $f_\epsilon$ is the $\nu F_\nu$ spectrum of 3C 454.3 measured at frequency $\nu = m_ec^2 \epsilon/h$,
and $E_1= m_e c^2\epsilon_1$ is the maximum photon energy during the period in which $f_\epsilon$ is measured.   The $\nu F_\nu$ flux $f_\epsilon$ in eq.\ (\ref{deltaD}) is evaluated at 
$\epsilon = \hat{\epsilon}={2\delta^2}/{(1+z)^2\epsilon_1}$ from the pair-production threshold condition.
Writing $f_{\epsilon} = 10^{-10}f_{-10}$ erg cm$^{-2}$ s$^{-1}$ in eq.\ (\ref{deltaD})
gives $\delta_{min}\approx 13\left[{f_{-10} E_1(10{\rm~GeV})}/{t_{{\rm var,d}}}\right]^{1/6}$
and $\hat{E}({\rm ~keV}) \cong 3.4 (\delta_{\rm min}/15)^2/E_1$(10 GeV).
Contemporaneous Swift XRT observations give the $\nu F_\nu$ 
flux between 0.2 and 10 keV, with
$\nu F_\nu(4 {\rm~keV}) \approx 6\times 10^{-11}$ erg
cm$^{-2}$ s$^{-1}$  \citep[e.g.,][]{Bon10}, so that $\delta_{\rm min}\approx 15$. 

The value of the Doppler factor can also be constrained from the argument that the energy flux from the Synchrotron Self-Compton (SSC) component should be below that observed at soft X-rays, in terms of isotropic luminosities $L_{\rm SSC}\le L_{\rm X}$. SSC luminosity is related to the synchrotron (SYN) luminosity via $L_{\rm SSC}/L_{\rm SYN} \sim u^\prime_{\rm SYN}/u^\prime_{\rm B}$ in the Thomson limit, where $u^\prime_{\rm SYN}\sim L^\prime_{\rm SYN}/(4\pi R^2c)$ is the energy density of synchrotron radiation and $u^\prime_{\rm B}=B^{\prime 2}/8\pi$ is the energy density of magnetic field. Noting that Lorentz transformation of bolometric luminosity is $L_{\rm SYN}=\delta^4 L_{\rm SYN}'$, we obtain
\begin{equation}
\delta^6 \ge \frac{2(1+z)^2L_{\rm SYN}^2}{L_{\rm X}c^3t_{\rm var}^2B^{\prime 2}}\,.
\end{equation}
This can be combined with a constraint from the external-Compton (EC) and synchrotron luminosities ratio,
\begin{equation}
\frac{L_{\rm EC}}{L_{\rm SYN}} \sim \delta^2\frac{u_{\rm EXT}}{u^\prime_{\rm B}}\,
\end{equation}
\citep{Der95}, where $u_{\rm EXT}$ is the energy density of external radiation. Eliminating from these equations the magnetic field value, we obtain
\begin{equation}
\delta \ge 17.6 \times L_{\rm SYN,48}^{1/8}\times L_{\rm EC,49}^{1/8}\times L_{\rm X,47}^{-1/8}\times t_{\rm var,d}^{-1/4}\times \left(\frac{u_{\rm EXT}}{0.015{\rm~erg~cm}^{-3}}\right)^{-1/8}\,.
\end{equation}
On the other hand, we can eliminate the Doppler factor and obtain a lower limit on the magnetic field
\begin{equation}
B^\prime \ge 3.4\;{\rm G} \times L_{\rm SYN,48}^{5/8}\times L_{\rm EC,49}^{-3/8}\times L_{\rm X,47}^{-1/8}\times t_{\rm var,d}^{-1/4}\times \left(\frac{u_{\rm EXT}}{0.015{\rm~erg~cm}^{-3}}\right)^{3/8}\,.
\end{equation}
Order-of-magnitude reference values of isotropic luminosities $L_{\rm SYN}$, $L_{\rm X}$ and $L_{\rm EC}$ have been deduced from Fig. 4 in \cite{Bon10}.
If the source region is located within the broad-line region (BLR), the energy density of external radiation is $u_{\rm EXT}\sim L_{\rm BLR}/(4\pi r_{\rm BLR}^2c)$, where $L_{\rm BLR}$ is the luminosity of the broad emission lines and $r_{\rm BLR}$ is the characteristic radius of the BLR. We approximate $L_{\rm BLR}$ with {\it GALEX} measurement of Ly$\alpha$ line, $L_{{\rm Ly}\;\alpha} \approx 2\times 10^{45}\;{\rm erg\;s^{-1}}$ and adopt $r_{\rm BLR} \approx 6\times 10^{17}\;{\rm cm}$ based on reverberation scaling relations for CIV line \citep{Bon10,Kas07}.

The analysis of radio observations made at a different epoch gives 
$\delta=24.6\pm 4.5$, bulk Lorentz factor $\Gamma = 15.6\pm 2.2$,
and observing angle $\theta = 1.3^\circ\pm 1.2^\circ$ 
obtained from superluminal observations \citep{Jor05}.
Both our constraints on the Doppler factor are consistent with this result and would become even tighter, if we adopt a shorter variability timescale.
Hence, we take $\delta_{25}\approx \Gamma_{20} \approx 1$ ($\Gamma= 20\Gamma_{20}$).
The constraint on the value of the magnetic field $B^\prime$ is a few times the equipartition magnetic field $ B^\prime_{\rm eq}({\rm G}) \approx 2\left(f_{\rm -10}\Lambda_2\right)^{2/7}/\left(\delta^{13/7}_{25} \nu^{1/7}_{\rm 13} t^{6/7}_{\rm var,d}\right)$, where $f_{\rm SYN,peak}=10^{-10}f_{-10}$ erg cm$^{-2}$ s$^{-1}$ is the observed synchrotron peak flux, $\nu_{\rm SYN,peak}=10^{13}\nu_{13}$ Hz is the synchrotron peak frequency and $\Lambda$ is the product of the bolometric correction factor and  ratio of energy contents in protons and electrons \cite[e.g.,][]{Fin08}.

For a conical geometry of opening angle $\theta_j\gtrsim R/r\sim 1/\Gamma$, the location of the emitting region for
the December flare  is constrained to be at distance
$r\lesssim 2c\Gamma^2t_{\rm var}/(1 + z)\approx 0.2\Gamma^2_{15}t_{\rm var,d}$ pc; i.e., towards
the outer parts of the BLR. A narrow jet with opening angle $\lesssim 0.2/\Gamma$, places the emitting region further out, and could be consistent with a fragmentation of outflow \citep{Mar10}. The most rapidly varying flaring episodes suggest
locations at the sub-pc scale, contrary to conclusions from coherent
optical polarization changes in 3C~279 \citep{2010Natur.463..919F} and PKS 1510-089 
\citep{2010arXiv1002.0806M} over timescales of tens of days,
which suggest that the emitting regions are several pc from the black hole.
Larger distances would be inferred if $\Gamma\gg 15$ or $\theta_j\ll 1/\Gamma$.
Higher $\Gamma$ factors would conflict with the expected source density of the parent population of FSRQs, as we discuss later. Narrow jets are suggested by radio observations \citep{2009A&A...507L..33P,Jor05} and could result from recollimation shocks at the pc scale \citep[e.g.,][]{2009MNRAS.392.1205N,2009ApJ...699.1274B} 
that reduce the characteristic size of the
emission region. A multiwavelength radio, mm, optical, and X-ray campaign covering observations of 3C~454.3 between 2005  and 2008 shows coherent optical and mm flux and polarization episodes consistent with the brightest events taking place at the mm core located $\sim$ pc beyond the acceleration and collimation zone \citep{Jor10}. Without strong collimation or recollimation, the short observed variability timescale implies a very small value $\theta_j\sim 2\times 10^{-3} t_{{\rm var,d}}/r_{pc}$ \citep{Tav10} with an associated low radiative efficiency.  Alternately, flaring episodes
with short variability times might take place within the BLR, whereas the
more slowly varying emissions could be radiated by jet plasma 
at larger distances. Spectral
variations due to the different target photon sources might be concealed by
mixing from components made at small and large radii. ``Residual" collisions
at large radii might also make a slowly varying underlying emission components,
as similarly inferred for GRB observations \citep{Li08}. 

The large changes in flux exhibited by the $\gamma$-ray light curves
in Fig.\ \ref{fig:light_curve} 
could be due to several different factors, including a changing
mean magnetic field, electron number, Doppler factor, target photon 
density and spectrum as the jet moves outward, or some combination of these factors. Any such 
explanation must also account for the moderate spectral changes, including
the near constancy of $E_{\rm break}$ (Fig.\ \ref{fig:Eb_flux}), while the flux changes by 
over an order of magnitude. 

One possibility is related to scattering of a target photon field in the Klein-Nishina regime. Compton scattering takes place in the Thomson limit when the energy of the photon to be scattered is (in $m_e c^2$ units) $\epsilon'' \lesssim 1/4$ in the electron rest-frame. Since the radiation from ultrarelativistic electrons is seen only when they are moving very close to the line of sight, that energy for $\gamma \gg \Gamma$ is $\epsilon'' \simeq \gamma \delta \epsilon_\star$, where $\epsilon_\star$
is the average energy of the target photons in the lab frame.
Hence, the scattering is in the Thomson regime provided
the upscattered photon energies are $\epsilon_C \simeq (4/3) \delta^2 \gamma^2 \epsilon_\star/(1+z) \lesssim 1/[12 \epsilon_\star(1+z)]$, i.e. for $E_C({\rm GeV}) \lesssim 12/E_\star({\rm eV})$,
independent of the Doppler factor \cite[compare][]{Geo01}.

\begin{figure}
\epsscale{0.6}
\plotone{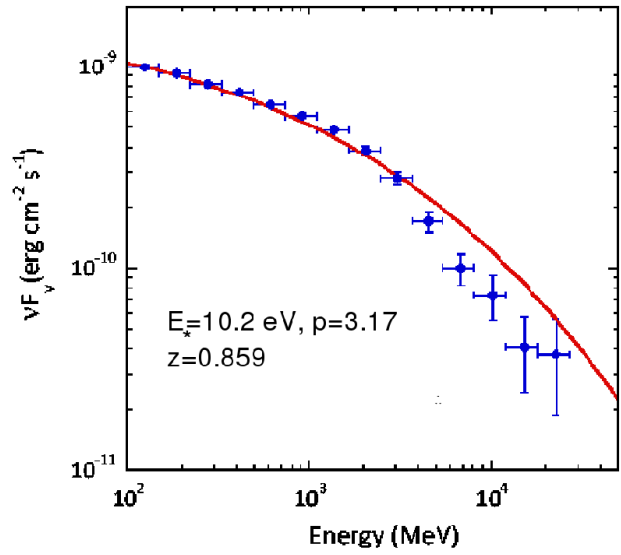}
\caption{ Model for the $\gamma$-ray spectrum of 3C 454.3 when 
a power-law electron distribution Compton scatters Ly $\alpha$ photons.
Best fit electron number index $p = 3.17$. Model is insensitive to values of 
lower and upper comoving electron Lorentz factors $\gamma_{min}$ and $\gamma_{max}$ 
provided $\gamma_{min} \lesssim 10^2$ and $\gamma_{max}\gtrsim 10^4$. The Klein-Nishina
softening from a power-law electron distribution gives a poor fit to the data.}
\label{fig:Lyalpha}
\end{figure}

If the break energy observed in 3C
454.3 at $\approx$ 2 GeV is due to the transition to scattering in
the KN regime, then the underlying target
photon energy $E_\star \approx 6\;{\rm eV}$ is
close to the energy of the Ly $\alpha$ photon at 10.2 eV. 
We have tested this possibility by comparing the Compton-scattered
spectrum from a power-law electron distribution with a monochromatic
Ly $\alpha$ photon source with the Fermi LAT data. 
The spectrum is independent of $\delta$, though can depend
on maximum and minimum electron Lorentz factors. As shown in Fig.\
\ref{fig:Lyalpha}, the spectrum is too hard to fit the data, and treatment of 
KN effects on cooling \citep{Der02, Mod05, Sik09}
or the addition of other soft photon sources will reduce the sharpness
of the break. The difficulty to fit the sharp spectral break with 
a single power-law electron distribution is in accord with the conclusion
of \citet{Abdo_3C} that this break reflects a complex electron 
spectrum.  To obtain a good spectral fit
to the $\gamma$-ray spectrum of 3C 454.3,
\cite{Fin10} use a broken power-law electron distribution and
a multi-component Compton-scattering model. The robustness of $E_{break}$ 
in this model is
due to similar, $\approx r^{-3}$ dependence of
accretion disk and external radiation energy density within the BLR.

Although \cite{Abdo_3C} excluded the possibility of explaining the spectral break with photon-photon pair absorption, \cite{2010arXiv1005.3792P} proposed that significant intrinsic absorption can be provided by He II Lyman-$\alpha$ line at $40.8\;{\rm eV}$ and continuum at $54.4\;{\rm eV}$. However their model implies the location of the gamma-ray production is in the inner edge of the BLR and inefficient scattering in the KN regime is predicted to lead to significant hardening of synchrotron spectra in IR/optical bands. Further studies are required to verify whether such effect is in contradiction  with multiwavelength spectra created within a one-zone model.

If $\Gamma \cong 15$ and the jet opening angle
$\theta_j\approx 1/\Gamma$, then the beaming factor for a two-sided top-hat
jet is $f_b\cong 1/2\Gamma^2\cong 1/(450\Gamma^2_{15})$. Within the framework of the 
unification hypothesis for radio galaxies \citep{Urr95}, 
FRII radio galaxies are the parent
population of FSRQs. The space density of FRII radio
galaxies at 0.8 $<$ z $<$1.5 is $\rho\cong 3 \times 10^{-7}$ Mpc$^{-3}$ 
\citep{Gen10}. At redshift unity, the volume
of the universe $\cong 4\pi/3R^3_H \cong 3 \times 10^{11}$ Mpc$^3$. Therefore,
if all FRII radio galaxies were misaligned FSRQs similar
to 3C~454.3, then the implied number $N$ of FSRQs is
$N < f_b \times 3 \times 10^{-7}$ Mpc$^{-3} \times 3 \times 10^{11}$ Mpc$^3 \approx 200/\Gamma^2_{15}$. 
This number is similar to the number, $\approx$ 100, of $\gamma$-ray
blazars at redshift unity \citep{1LAC}, so is in accord with the unification scenario.
This simple argument suggests that unless $\theta_j \gg 1/\Gamma$, 
in which case the total jet
energy of an individual FSRQ blazar would be correspondingly larger,
the typical bulk Lorentz factor $\Gamma$ for FSRQs would not be $\gg 15$.

\section{Summary}

The correlated spectral and temporal properties of 3C~454.3 during two very strong flaring episodes, during which
 the source was the brightest object in the $\gamma$-ray sky, have been studied.  The $\gamma$-ray spectrum of 3C 454.3 shows a significant spectral break between $\approx 2$ -- 3 GeV that is very weakly dependent on the flux state, even when the flux changes by an order of magnitude. The light curves during periods of enhanced activity in  July -- August 2008 and December 2010 show strong resemblance, with a flux plateau of a few days preceding the major flare. The spectral index measured on a daily basis shows a very moderate ``harder when brighter" effect for fluxes measured at $E>E_1$, where $E_1 = 163$ MeV is chosen to minimize spurious correlations resulting from the choice of the low-energy bound on the range.  However, the weekly spectral index displays a more significant variation,  but not exceeding $\Delta\Gamma$=0.35 when the flux varies by more than a factor of 10.  Indication for a gradual hardening preceding a major flare and extending for several weeks is found.  No recurring pattern in the photon spectral index/flux plane measured at $E>E_1$ during the course of flaring episodes has been identified. Flux variations of a factor of 2 have been observed over time scales as short as a few hours, though only weak variability was observed during the time of the Target of Opportunity pointing of the Fermi Telescope towards 3C 454.3.

From $\gamma\gamma$ opacity constraints, we derive a minimum Doppler factor  $\delta_{min}\cong 15$ from the flux, variability time, and highest energy photon measurements. This value is in accord with independent measurements of $\delta$ from superluminal motion observations \citep{Jor05}. The behavior of the break energy has also been investigated. Spectral softening due to the onset of Klein-Nishina effects in Compton scattering, which is independent of Doppler factor for a power-law electron distribution, was considered as the origin of the $\gamma$-ray spectrum. The magnitude of the softening for such a model was found to be inadequate to fit the spectrum without introducing intrinsic breaks in the underlying electron spectrum. An independent estimate of $\delta$ and $B$ also gives values of $\delta \gtrsim 15$; also $B \gtrsim$ several G. The short flaring times suggest an origin of the emission region within the BLR, but uncertainties in parameters are compatible with the production of emission from 3C 454.3 on the pc scale.    

\acknowledgments The \textit{Fermi} LAT Collaboration acknowledges generous
ongoing support from a number of agencies and institutes that have supported
both the development and the operation of the LAT as well as scientific data
analysis.  These include the National Aeronautics and Space Administration and
the Department of Energy in the United States, the Commissariat \`a l'Energie
Atomique and the Centre National de la Recherche Scientifique / Institut
National de Physique Nucl\'eaire et de Physique des Particules in France, the
Agenzia Spaziale Italiana and the Istituto Nazionale di Fisica Nucleare in
Italy, the Ministry of Education, Culture, Sports, Science and Technology
(MEXT), High Energy Accelerator Research Organization (KEK) and Japan
Aerospace Exploration Agency (JAXA) in Japan, and the K.~A.~Wallenberg
Foundation, the Swedish Research Council and the Swedish National Space Board
in Sweden. Additional support for science analysis during the operations phase is
gratefully acknowledged from the Istituto Nazionale di Astrofisica in Italy and
the Centre National d'\'Etudes Spatiales in France.


\bibliography{3C454_Dec09.bib}
\clearpage


\clearpage

\clearpage

\clearpage

\clearpage

\clearpage

{\footnotesize 
\begin{table}
\noindent \begin{tabular}{|c|c|c|c|c|c|c|c|}
\hline
period & $\Gamma_1$ & $\Gamma_2$ &  E$_{break}$  & $\alpha$ & $\beta$ & $\Gamma$ & E$_{cutoff}$  \\
& & & (MeV) & & & & (MeV) \\
\hline
1 & 2.31$\pm$ 0.02 & 3.19$\pm$0.12 & 1810$_{-220}^{+250}$ & 2.52$\pm$0.03 &  0.11$\pm$0.02 & $2.29\pm0.03$ & 5200$\pm$900 \\
\hline
2 & 2.33$\pm$ 0.03 & 3.29$\pm$0.21 & 2750$_{-360}^{+470}$ &  2.48$\pm$0.03 & 0.09$\pm$0.02 & 2.25$\pm0.04$ & 7000$\pm$1700  \\
\hline
3 & 2.32$\pm$ 0.02 & 3.12$\pm$0.08 & 1630$_{-160}^{+180}$ &  2.54$\pm$0.02 & 0.12$\pm$0.01 & 2.24$\pm0.03$ & 5100$\pm$620  \\
\hline
4 & 2.38$\pm$ 0.02 & 3.23$\pm$0.14 & 2380$_{-280}^{+750}$ &  2.55$\pm$0.03 & 0.10$\pm$0.02 & 2.30$\pm0.03$ & 7000$\pm$1300  \\
\hline
\end{tabular}
\caption{Parameters of the functions fitted to the spectra for the different periods considered in Figure \ref{fig:nuFnu}.}
\label{tab:funct}
\end{table}
\end{document}